\documentclass[fleqn,twoside,fps]{article}
\usepackage{espcrc2,graphicx,latexsym}
\usepackage[figuresright]{rotating}
\usepackage{bm}
\usepackage{color}
\usepackage{fancybox}
\usepackage{amsmath}
\usepackage[psamsfonts]{amssymb}
\usepackage{rsfs}

\title{Gluon and ghost propagators   
from the viewpoint of 
general principles of 
quantized gauge field theories  
}

\author{Kei-Ichi Kondo\address[Wup]{Department of Physics, Faculty of Science, 
Chiba University, Chiba 263-8522, Japan}
        \addressmark[Oxf]\thanks{Supported by 
Sumitomo Foundations, Grant-in-Aid for Scientific Research  (B)13135203 from MEXT,  and  (C)14540243  from JSPS.
}}

\begin{document}

\begin{abstract}
We discuss the possible form of gluon and ghost propagators in the infrared region of Yang-Mills theory in the covariant gauge from the viewpoint of general principles of 
quantized gauge field theories.  
\end{abstract}

\maketitle

\section{INTRODUCTION}
Recent studies of the coupled Schwinger-Dyson (SD) equations for Yang-Mills theory have shown \cite{AS01} that  the 
gluon  propagator $D_T$ and  ghost propagator $\Delta$ in the Landau gauge exhibit {\it the power law behavior with a critical exponent $\kappa$}
\begin{equation}
  D_T(Q^2) :=F(Q^2)/Q^2 \cong A(Q^2)^{2 \kappa -1}, 
\end{equation}
\begin{equation}
  \Delta(Q^2) := G(Q^2)/Q^2 \cong B(Q^2)^{- \kappa -1}  ,
\end{equation}
in the infrared (IR) limit $Q^2 \rightarrow 0$ for Euclidean momentum $Q^2>0$.
Surprisingly, {\it the gluon propagator is IR suppressed}, while 
{\it the ghost propagator is IR enhanced} for $\kappa \ge 1/2$.  
However, 
\begin{enumerate}
\setlength{\itemsep}{-3pt}
\item The {\it precise value of $\kappa$} is unknown even if the IR power law is correct.  The transverse gluon propagator vanishes for $\kappa >1/2$, while it has a non-zero limit for 
 $\kappa=1/2$.
\item There is no {\it analytical}  method to connect the IR asymptotic solution to the UV one, although numerical methods exist. 
There is no guarantee for 
 uniqueness of the solution obtained under the specific Ansatz.  
\item  There is no argument for the analytic continuation from Euclidean region to Minkowski region.  
\end{enumerate}

The purpose of this work \cite{Kondo03} is to search the solution which is consistent with the  general principles  of quantized gauge field theory:
\begin{itemize}
\setlength{\itemsep}{-3pt}
\item Non-perturbative multiplicative renormalizability  
\item Analyticity 
\item Spectral condition 
\item Poincar\'e group structure 
\end{itemize}
 without using 
 the  SD equations.

\section{Multiplicative renormalizability}

Non-perturbative multiplicative renormalizability for the 
 gluon form factor, 
$
  F_0(k^2,\Lambda^2,\alpha_0,\lambda_0) 
$
$  = Z_3(\mu^2,\Lambda^2) F_R(k^2,\mu^2,\alpha,\lambda) ,
$
yields the RG equation:
\begin{multline}
\left[ \mu {\partial \over \partial \mu} + \beta(\alpha) {\partial \over \partial g^2} 
- 2 \lambda \gamma_{\lambda}(\alpha) {\partial \over \partial \lambda} 
+ 2 \gamma_{\mathscr{A}}(\alpha)   \right]
\\
	\times F_R \left({k^2 \over \mu^2},\alpha,\lambda \right)  = 0 . \end{multline}
where
$\alpha:=g^2/(4\pi)$, 
$
  \beta(\alpha) := \mu {\partial \alpha(\mu) \over \partial \mu} ,
$
$
 \gamma_{\mathscr{A}}(\alpha) := {1 \over 2} \mu {\partial \ln Z_3(\mu) \over \partial \mu} ,
$
and 
$
   \gamma_{\lambda}(\alpha)  
:= {1 \over 2}  \mu  {\partial \ln Z_{\lambda} \over \partial \mu}
$ 
for a gauge parameter $\lambda$.


\begin{figure}[htbp]
\begin{center}
\includegraphics[height=3.5cm]{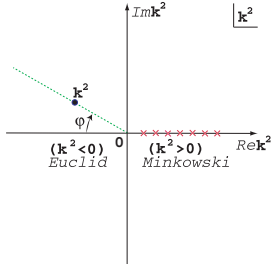}
\caption{\small The ray with angle $\varphi$ measured from the negative real axis $k^2<0$ on the complex $k^2$ plane with singularities on the positive real axis.}
\label{fig:ray}
\end{center}
\end{figure}


We consider the propagator along the ray in the cut $k^2$ plane (See Figure~\ref{fig:ray}).
The asymptotic freedom (i.e., validity of perturbation theory for large $k^2$) and the general solution of the RG equation 
lead to 
the asymptotic form for $|k^2| \rightarrow \infty$ along the ray:
$
 F(k^2) \cong  
 C_\mathscr{A}(\lambda) \left( \ln {|k^2| \over |\mu^2|} \right)^{-\gamma^\mathscr{A}_0(\lambda)/\beta_0} ,
$
$
 D_T(k^2) \cong   
- C_\mathscr{A}(\lambda) k^{-2} \left( \ln {|k^2| \over |\mu^2|} \right)^{-\gamma^\mathscr{A}_0(\lambda)/\beta_0} ,
$
where
$
 C_\mathscr{A}(\lambda)   > 0
$
with 
$  
 \beta(\alpha) = -{\beta_0 \over 2\pi} \alpha^2 + \cdots, 
$
 and 
$
 \gamma_\mathscr{A}(\alpha) = - {\gamma_0^\mathscr{A} \over 4\pi} \alpha + \cdots$. {\it 
Both functions $F(k^2)$ and $D_T(k^2)$ vanish as $|k^2| \rightarrow \infty$ along the ray}, since $\gamma^\mathscr{A}_0(\lambda)/\beta_0=13/22>0$.

\section{Analyticity}

\begin{figure}[htbp]
\begin{center}
\includegraphics[height=3.5cm]{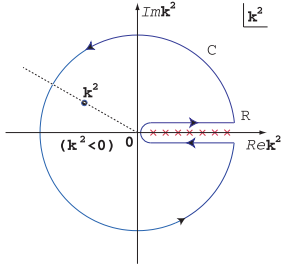}
\caption{An integration contour $C$ on the complex $k^2$ plane with singularities on the positive real axis.}
\label{fig:CutComplexPlane}
\end{center}
\end{figure}


Suppose 
i) the complex function $f(z)$ is {\it analytic} in the whole complex  plane $z:=k^2$ except for the positive real axis $z > 0$,
ii)    $f(z)$ {\it vanishes asymptotically} along any ray in the cut complex plane: $|f(z)| \rightarrow 0$ as $|z| \rightarrow \infty$,
and  
iii) $f(k^2)$ is {\it real on the real axis} $k^2<s_{min}$ (at least for the space-like region $k^2<0$).
Then, for a reference point $k^2$, 
 the {\it dispersion relation} follows  
\begin{equation}
  f(k^2) =  {1 \over \pi} \int_{0}^{\infty} dz {Im f(z+i\epsilon) \over z-k^2} .
\end{equation}
We apply this relation to the form factor and the propagator.

\section{Spectral condition}
From the assumptions:
1) Poincare group structure (representation),
2) spectral condition,
and
3) completeness condition,
the spectral representation for the 
renormalized gluon propagator follows:
\begin{equation}
  D_T(k^2) = \int_{0}^{\infty} dp^2 {\rho(p^2) \over p^2-k^2}, 
\end{equation}
where
$
  \rho(k^2) =  \pi^{-1} Im D_T(k^2+i\epsilon) .
$
 The spectral weight function   
$
  \pi \rho(k^2) = {\rm Im} D_T(k^2+i\epsilon) 
 $
 $
  = -k^{-2} {\rm Im} F(k^2+i\epsilon) , 
  $
  for 
  $
 k^2 > 0 ,
$
has the asymptotic behavior:
$
\rho_{as} \cong  -4\pi (k^2)^{-1} \gamma^\Phi_0(\lambda) |\beta_0^{-1}| C_\Phi(\lambda) \left( \ln { k^2 \over |\mu^2| } \right)^{-\gamma^\Phi_0(\lambda)/\beta_0-1} 
$
for $k^2 \rightarrow \infty $.
Here  the anomalous dimension $\gamma^\mathscr{A}$ is gauge  dependent.  
The sign of the asymptotic discontinuity $\rho_{as}$ is determined by the ratio $\gamma_0/\beta_0$, it is negative for $\gamma_0/\beta_0>0$.

\section{Superconvergence for gluon}
For $\gamma^A_0/\beta_0>0$, 
the {\it gluon propagator} has an unsubtracted {\it renormalized} dispersion relation 
\begin{equation}
  D_T(k^2 ) 
 = \int_{0}^{\infty} dp^2 {\rho(p^2 ) \over p^2 -k^2} .
\end{equation}
The renormalized $\rho$ is a function of $p^2, \mu, g_R, \lambda_R$ where $\mu$ is the renormalization point and $\lambda$ is the gauge parameter.
In the similar way, the {\it gluon dressed function or form factor} $F=-k^2 D_T$ has the {\it renormalized} dispersion relation 
\begin{equation}
  F(k^2 ) = {\lambda \over \lambda_*} - \int_{0}^{\infty} dp^2  {p^2 \rho(p^2 ) \over p^2 -k^2}    .
\end{equation}
Two relations are compatible if and only if  
\begin{equation}
  \int_{0}^{\infty} dk^2 \rho(k^2, \mu^2, g_R, \lambda_R) = {\lambda \over \lambda_*}  .
\end{equation}
This is the {\it superconvergence relation \cite{OZ80a} in the  generalized Lorentz gauge} with an {\it initial} gauge-fixing parameter $\lambda$ and the fixed point $\lambda_*$.

 We put the reference point $k^2$ on the Euclidean region, i.e., $Q^2=-k^2>0$. 
{\it The gluon form factor vanishes in the Euclidean IR limit $Q^2 \downarrow 0$}:
\begin{equation}
  F(0 ) 
 = {\lambda \over \lambda_*} - \int_{0}^{\infty} dp^2  \rho(p^2 ) 
= 0 ,
\end{equation}
and {\it the gluon propagator converges to a constant}
\begin{equation}
  D_T(-Q^2 ) 
=  {\it \int_{0}^{\infty} dp^2 {\rho(p^2 ) \over p^2} }
+ O(Q^2)  .
\end{equation}

The spectral function has the form 
$
  \rho(s) = Z \delta(s-M^2) + \tilde{\rho}(s)      
$
where $Z \delta(s-M^2)$ corresponds to a pole at $p^2=M^2$ and $\tilde{\rho}(s)$ is the contribution of the continuous spectrum from more than two particle states, 
namely, $\text{supp}\rho(s) \in [(2M)^2,\infty)$.
 The {\it finiteness of the IR limit $D_T(0)$} given by
\begin{equation}
  D_T(0) 
  = {Z \over M^2} + \int_{(2M)^2}^{\infty} ds \ s^{-1} \tilde{\rho}(s)
\end{equation}
is interpreted as indicating the {\it existence of a massive pole} ($M\not=0$), provided that $Z\not=0$. (The last integral is finite for $\gamma^A_0/\beta_0>0$.)
Thus, we have shown \cite{Kondo03} that, {\it for  gluon with  massive spectrum, the power-series expansions  for the gluon propagator and the form factor can be well-defined for small Euclidean momenta $Q^2$}
for any gauge parameter.
Note that $0< D_T(0) < \infty$ corresponds to $\kappa=1/2$. 
(The Gribov limit is $\kappa=1$.)

\section{Superconvergence for ghost?}

\begin{figure}[htbp]
\begin{center}
\includegraphics[height=3.5cm]{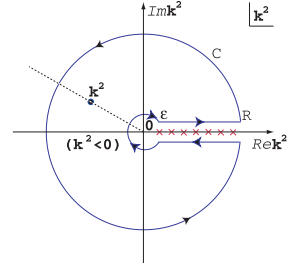}
\caption{The integration contour $C$ to which the Cauchy theorem is applied on the complex $k^2$ plane with singularities on the positive real axis.}
\vspace{-0.5cm}
\label{fig:MomentContour}
\end{center}
\end{figure}


We recall the result of \cite{Xu96}.
The {\it renormalized} ghost propagator $\Delta_{FP}$ has a dispersion relation in the arbitrary gauge. 
However, 
the form factor $G:=-k^2 \Delta_{FP}$ has a dispersion relation  
only in the Landau gauge $\lambda=0$.
Hence, {\it the superconvegence relation for the ghost holds,
$  
 0 =  \int_{0}^{\infty} dk^2 \rho_{FP}(k^2, \mu^2, g_R^2, 0) , 
$
only in the Landau gauge $\lambda=0$.
}
{\it 
For $\lambda\not=0$, however, the superconvergence relation does not hold}, since  the unsubtracted dispersion relation exists only for the propagator, not for the form factor, in sharp contrast with the gluon case. 

If the spectral function $\rho$ has singularities accumulating toward the origin $p^2=0$, however, we must replace an integration contour in Figure~\ref{fig:CutComplexPlane} by
another contour in Figure~\ref{fig:MomentContour} to avoid the origin.
We conclude that {\it the superconvergence for ghost does not hold even in the Landau gauge} 
 \cite{Kondo03}.

\section{CONCLUSIONS} 
From the viewpoint of general principles of QFT, 
{\it the transverse gluon can be massive and short range, while the FP ghost is singular and long-range}. 
 The {\it IR critical exponent of gluon is $\kappa=1/2$}, 
since the gluon propagator behaves like 
\begin{equation}
 D_T(Q^2) \cong \text{const.}   + O(Q^2) .
\end{equation}
 {\it The ghost propagator has  a negative and non-integer exponent.}    
It   fulfills a sufficient condition for  {\it color confinement} due to  Kugo and Ojima.
\begin{equation}
 \lim_{Q^2 \rightarrow 0} [Q^2 \Delta(Q^2)]^{-1} 
\equiv  \lim_{Q^2 \rightarrow 0} [G(Q^2)]^{-1}
= 0 .
\label{KO}
\end{equation}
 Supposing the {\it existence of IR fixed point} for the gluon--ghost--antighost coupling constant (without dynamical quarks), as suggested from the SD equations \cite{AS01}, we have
\begin{equation}
 \Delta(Q^2) \cong (Q^2)^{-3/2} . 
\end{equation}
 Thus the ghost is expected to be a carrier of confinement in the covariant gauge.

\end{document}